\documentclass[12pt]{article}
\usepackage{latexsym, amsfonts, amsmath}

\usepackage[active]{srcltx}
\usepackage{epsfig}
\usepackage{graphicx}

\DeclareMathOperator{\sign}{sign}

\newtheorem{thm}{Theorem}[section]

\newtheorem{exam}[thm]{Example}

\def\am{{\rm argmax\,}}

\begin{document}

\bibliographystyle{plain}

\title {Strength of Forensic Evidence
for Composite   Hypotheses:\\ An   Empirical Bayes View with a Fixed Prior Quantile}

\author {Bert van Es\\[0.3cm]
{\normalsize Korteweg-de Vries Institute for Mathematics}\\
{\normalsize University of Amsterdam}\\
{\normalsize Science Park 105-107,
 1098GX Amsterdam}\\
{\normalsize The Netherlands,}\\
{\normalsize Email: A.J.vanEs@uva.nl}}


\maketitle

\begin{abstract}
Motivated by the forensic problem of determining the strength of evidence of a continuously distributed measurement of evidence, in the situation of composite hypotheses of the prosecutor and the defence concerning a parameter of a parametric model, we consider empirical Bayes methods with a prescribed quantile value for the prior distribution.

Firstly we derive the strength of evidence for nonparametric priors. It turns out that we get the by now more or less accepted strength of evidence as the ratio of two suprema, $\sup_{\theta\geq\theta_0}f(x|\theta)/\sup_{\theta<\theta_0}f(x|\theta)$. Here the hypotheses of the prosecutor and defence are given by $H_p: \theta\geq \theta_0$ and $H_d:\theta<\theta_0$. The evidence is seen as a measurement $x$ which is a realization of a random variable with a density $f(x|\theta)$.

Secondly we consider a similar parametric empirical Bayes method with a quantile restriction on the prior where the prior distribution is assumed to be normal. Some interesting strength of evidence functions are derived for this situation.
\\[.5cm]

\end{abstract}

\section{Introduction}

In forensic science the key persons in a trial are a suspect, a prosecutor,
a defence attorney and a judge.
The judge has to rule about several questions: did the crime actually happen?, is it illegal?, did the suspect commit the crime?, is the suspect guilty? is the suspect punishable? and what should the sentence be? We focus on  the guilt of the suspect  and we only consider one particular piece of evidence. We will focuss on the strength of this evidence.

The strength of a certain piece of evidence $E$ is nowadays often interpreted
in the context of the Bayesian paradigm. The prosecutor and the defense attorney
will have two different hypotheses corresponding to the evidence, denoted by $H_p$ and $H_d$. We  will assume that these hypotheses are mutually exclusive.
For instance the prosecutor will say that the suspect left the blood trace with its specific DNA profile at the crime scene, and the defence
attorney will say that an arbitrary other person left the trace.
The fundamental equation of the Bayesian paradigm of forensic statistics for discrete types of evidence such as DNA profiles, shoe sole patterns etc. states
\begin{equation}\label{bp}
{P(H_p|E)\over P(H_d|E)}={P(E|H_p)\over P(E|H_d)}{P(H_p)\over P(H_d)}
\end{equation}
which follows from the Bayes rule in the odds formulation.

According to the Bayesian paradigm the judge is supposed to formulate his or her prior odds
concerning $H_p$ and $H_d$.
By the Bayes rule the posterior odds, i.e. the odds given the evidence,
are then equal to the product of the likelihood ratio, the first term on the right hand side of (\ref{bp}), and the prior odds, the second term of (\ref{bp}).
So we define
\begin{equation}\label{lr}
V={P(E|H_p)\over P(E|H_d)}.
\end{equation}
This likelihood ratio $V$ is called the {\em  strength, of the evidence}. In general statistics it is called the Bayes factor, see for instance Kass and Raftery (1995). Values larger than one support the prosecutors hypothesis and values lower than one support the defense hypothesis.

An important example  where this paradigm is applicable is for instance  blood trace DNA (as a set of observed DNA features). It is also applied to many other types of discrete evidence. Note that more involved DNA analyses also may include peak heights which are then analysed as continuous evidence. For a more detailed description of the Bayesian paradigm we refer to Aitken and Taroni (2004).

An important observation here is that in simple discrete situations as above the likelihood ratio does not depend on the prior odds. This means that it
quantifies the strength of the evidence, whatever the judge decides to choose as prior odds. So its value is inherent to the evidence alone and it can be reported by a forensic expert without knowing the prior odds.

\bigskip

In cases, where for instance the blood alcohol content of the blood of a suspect is measured with a measurement error,
the situation is quite different. The evidence is then a measurement $x$, the blood alcohol content in a blood sample from a suspect.
The prosecutor will say that the true blood alcohol level is above some legally allowed level $\theta_0$, and
the defense attorney will say that it is not. Hence if we model the measurement $x$ as a realization of a random variable with a density
$f(x|\theta)$, where $\theta$ denotes the true unknown level, then the hypotheses will be of the form $H_p: \theta\geq \theta_0$ and $H_d:\theta<\theta_0$. The judge
has to assign prior probabilities to these. This means that we have to adapt a more involved approach to the problem and assume that the parameter $\theta$ has a prior distribution.

We will discuss the strength of evidence in this context. We will derive the by now
more or less accepted value of evidence as the ratio of two suprema,
\begin{equation}\label{supsratio}
V(x)=\frac{\sup_{\theta\geq\theta_0}f(x|\theta)}{\sup_{\theta<\theta_0}f(x|\theta)},
\end{equation}
similar to the likelihood ratio statistic, from a nonparametric empirical Bayes point of view. Subsequently we will describe the somewhat surprising results of a parametric empirical Bayes approach, where we assume a parametric normal model for the prior density, to this problem. For a different motivation of this value of evidence formula see Bickel (2012).
\\\\\\

\section{Continuous models with composite hypotheses}

For continuously distributed evidence the strength of evidence is taken to be equal to the ratio of the densities under both hypotheses, instead of the ratio of two probabilities.
We then get
$$
V(x)={f(x|H_p)\over f(x|H_d)}.
$$
Here the realization $x$ is seen as the evidence. An example is for instance a height measurement $x$ of a person on a vague CCTV film of a robbery of a store.
Relevant hypotheses could be $H_p:$ "the perpetrator is an arbitrary man from a certain reference population" and $H_d:$ "the perpetrator is an arbitrary
woman from that population".

The above example is essentially different from the situation where we pose a distribution from a parametric family
for the evidence and where the hypotheses concern the parameter.
Let us suppose that we have to determine the strength of the evidence of a measurement $x$ which we
can see a realization of a density $f(x|\theta)$ belonging to a parametric family.
Recall that we write $H_p$ for the hypothesis of the prosecutor and $H_d$ for the hypothesis of the defence.
Suppose that these two hypotheses can be expressed in terms of the parameter $\theta$ as
\begin{equation}\label{hyps}
H_p: \theta\geq \theta_0\quad\mbox{and}\quad
H_d: \theta< \theta_0.
\end{equation}
The judge expresses his or her prior belief in the hypotheses in terms of the probabilities $P(H_p)$ and $P(H_d)$.

If we adopt the frequentist point of view  the probabilities $P(H_p)$ and $P(H_d)$ have no meaning because the parameter is fixed and has no probability distribution. This can be solved by adding an underlying    Bayesian  prior distribution for the parameter $\theta$.
Let us assume that we have a prior distribution $\pi(\theta)$ for $\theta$.
In our notation we will consider $\pi$ to be a density of a continuous distribution but in fact we impose no such restrictions. Discrete priors $\pi$ are also allowed.
This prior has to satisfy the prior probabilities posed
by the judge. So the probabilities
\begin{equation}\label{restrictions}
P(H_p)=\int_{\theta_0}^\infty   \pi(\theta)d\theta
\quad\mbox{and}\quad
P(H_d)=\int_{-\infty}^{\theta_0} \pi(\theta)d\theta
\end{equation}
are fixed by the prior belief of the judge. Note that we explicitly require the judge to give only the probabilities of the hypotheses and that we consider all prior distributions that are coherent with these probabilities.
If we see the strength of evidence as the ratio of the posterior odds and the prior odds (the Bayes Factor), as is evident in (\ref{bp}), then
 for continuously distributed evidence we get,
for a given underlying prior $\pi(\theta)$  for $\theta$,
\begin{equation}\label{voe}
V(x)={ \int_{\theta_0}^\infty f(x|\theta)\pi(\theta)d\theta/ P(H_p)\over \int_{-\infty}^{\theta_0}f(x|\theta)\pi(\theta)d\theta/P(H_d)}.
\end{equation}
This strength depends on the   prior $\pi$ of the judge which is now not completely determined by the
prior belief in $H_p$ and $H_d$ of the judge. In fact only the $\theta_0$ quantile of the distribution $\pi$ is fixed by (\ref{restrictions}). For the rest it
is arbitrary and possibly discrete.
If the judge would provide the full prior $\pi$, implicitly fixing the probabilities $P(H_p)$ and $P(H_d)$, then the strength of evidence would be given by (\ref{voe}) for his or her specific prior.

\bigskip

\begin{exam} {Blood alcohol measurements.}

{\rm
Our running example will be blood alcohol measurements, see Taroni, Biedermann, Bozza,  Vuille and  Augsburger (2014) for a review. A driver has been apprehended on suspicion of having drunk too much alcohol. A blood test has been performed and $x$ permille alcohol has been measured in his or her blood. We will see this value $x$ as the evidence. Let assume that the legally permitted permillage is $\theta_0$.
The statistical model assumes that the true permillage is $\theta$ and that the measurement device has a normally distributed error with mean zero and standard error $\sigma$, which is assumed to be known. Hence the measurement
$X$ has a ${\cal N}(\theta,\sigma^2)$ distribution. So, with $\phi$ denoting the standard normal density we have
\begin{equation}\label{normalba}
f(x|\theta)=\frac{1}{\sigma}\phi\Big(\frac{x-\theta}{\sigma}\Big).
\end{equation}
The hypotheses of the prosecution and the defence in this situation are given by (\ref{hyps}).
}
\end{exam}

\section{Nonparametric empirical Bayes}\label{nonparametric}

Let us use a nonparametric  empirical Bayes approach. We will use our one observation $x$ to estimate the prior density $\pi$ by maximum likelihood
under the fixed quantile restriction (\ref{restrictions}) posed by the judge. Let us write $V_{npar}$ for the resulting estimate of the strength of evidence
(\ref{voe}).

\begin{thm}
Using the nonparametric empirical Bayes approach with a maximum likelihood estimate of the prior distribution $\pi$ we have
\begin{equation}\label{supsratio2}
V_{npar}(x)=\frac{\sup_{\theta\geq\theta_0}f(x|\theta)}{\sup_{\theta<\theta_0}f(x|\theta)}.
\end{equation}
\end{thm}

\noindent{\em Proof}

Note that the density of the evidence $x$ is given by
\begin{equation}\label{g}
g(x)=\int_{-\infty}^\infty f(x|\theta)\pi(\theta)d\theta.
\end{equation}
If we maximize this value, the likelihood of the evidence, over all priors $\pi$ which satisfy the belief of the
judge stated in (\ref{restrictions}), then we get that the maximizing distribution $\pi_{emp}$ is a discrete distribution with two
values $\theta_p$ and $\theta_d$, given by
$$ \theta_p=\am_{\theta\geq\theta_0} f(x|\theta)\quad\mbox{and}\quad
\theta_d=\am_{\theta<\theta_0} f(x|\theta),
$$
which are attained with probabilities $P(H_p)$ and $P(H_d)$ respectively.
Since $P(H_p)+P(H_d)=1$ all the mass of this prior in concentrated in the points $\theta_p$ and $\theta_d$.
Note that this prior depends on the evidence, just like empirical Bayes priors depend on the data.
For this empirical prior the strength of the evidence (\ref{voe}) becomes
$$
{ \int_{\theta_0}^\infty f(x|\theta)\pi_{emp}(\theta)d\theta/ P(H_p)\over \int_{-\infty}^{\theta_0}f(x|\theta)\pi_{emp}(\theta)d\theta/P(H_d)}
={{f(x|\theta_p)}\over{f(x|\theta_p)}}
={\sup_{\theta\geq\theta_0}f(x|\theta) \over{\sup_{\theta<\theta_0}f(x|\theta)}}.
$$

\hfill $\Box$

An interesting application of this strength of evidence representation to scientific integrity studies can be found in Klaassen (2015).

\bigskip
\begin{exam}{Blood alcohol measurements continued.}

{\rm If we determine the strength of evidence in our example of  blood alcohol measurements, where we have density (\ref{normal}) for our evidence, then we get
\begin{equation}\label{voenpar}
V_{npar}(x)=
\left\{
\begin{array}{ll}
e^{-\frac{1}{2}\, \frac{(x-\theta_0)^2}{\sigma^2}}
&, \mbox{if}\quad x\leq\theta_0,\\
e^{\frac{1}{2}\, \frac{(x-\theta_0)^2}{\sigma^2}}&, \mbox{if}\quad x\geq\theta_0.
\end{array}
\right.
\end{equation}
This strength of evidence function is given in Figure \ref{nparblood}.
\begin{figure}[h]
 \centering
  \includegraphics[height=2in,width=2.5in]{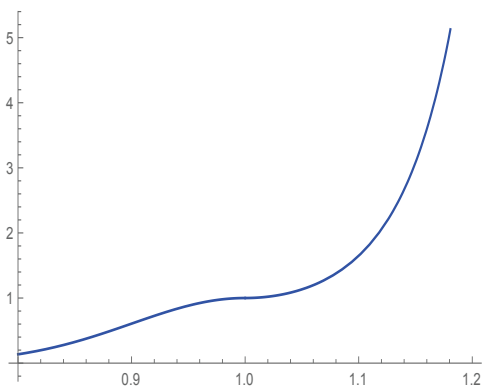}\\
  \caption{The function $V_{npar}$ for $\theta_0=1$ and $\sigma=0.1$.}\label{nparblood}
\end{figure}
}
\end{exam}

\begin{exam}{Strength of evidence functions for location and scale models.}

{\rm The blood alcohol example is an example of a specific model where the family of densities $f(x|\theta)$ is a {\em location family}, i.e. we have
\begin{equation}
f(x|\theta)=k(x-\theta),
\end{equation}
for some fixed density function $k$ and $-\infty<\theta<\infty$, which for convenience we assume to be strictly positive.
In this case we have
$$
\sup_{\theta\geq\theta_0}f(x|\theta)=\sup_{\theta\geq\theta_0}k(x-\theta)=\sup_{t\leq x-\theta_0}k(t).
$$
Hence we see that the numerator in (\ref{supsratio2}) is non decreasing. Similarly it follows that for negative $x$  the denominator in (\ref{supsratio2})
is non increasing. This shows that for a location family the nonparametric strength of evidence function is always non decreasing.

If $k$ is unimodal and symmetric around zero then the value at $x=\theta_0$ equals one.
If $k$ is multi modal then it turns out that the strength of evidence function has flat parts.

Let us, as a side step, now also consider the admittedly forensically less relevant {\em scale families} where
\begin{equation}
f(x|\theta)=\frac{1}{\theta}\,k\Big(\frac{x}{\theta}\Big),
\end{equation}
for some fixed density function $k$ and and $\theta >0$. For convenience we assume that $k$ is positive on the real line.
For these families we have, for positive $x$ and $\theta_0>0$,
$$
\sup_{\theta\geq\theta_0}f(x|\theta)=\sup_{\theta\geq\theta_0}\frac{1}{\theta}\,k\Big(\frac{x}{\theta}\Big)=\frac{1}{x}\,\sup_{0<t\leq x/\theta_0} tk(t)\ \mbox{and}\
\sup_{\theta<\theta_0}f(x|\theta)=\sup_{\theta\geq\theta_0}\frac{1}{\theta}\,k\Big(\frac{x}{\theta}\Big)=\frac{1}{x}\,\sup_{t>x/\theta_0} tk(t).
$$
Similarly for negative $x$ it follows that
$$
\sup_{\theta\geq\theta_0}f(x|\theta)=\sup_{\theta\geq\theta_0}\frac{1}{\theta}\,k\Big(\frac{x}{\theta}\Big)=\frac{1}{x}\,\inf_{ x/\theta_0\leq t<0} tk(t)\ \mbox{and}\
\sup_{\theta<\theta_0}f(x|\theta)=\sup_{\theta<\theta_0}\frac{1}{\theta}\,k\Big(\frac{x}{\theta}\Big)=\frac{1}{x}\,\inf_{t<x/\theta_0} tk(t).
$$
For $k=\phi$, the standard normal density, the function $t\phi(t)$ has a negative minimum at $t=-1$, it is zero at $t=0$ and it has  a positive maximum at $t=1$. In this example we have
$$
V_{npar}(x)=
\left\{
\begin{array}{ll}
\frac{|x|}{\theta_0}\,\frac{\phi(\frac{|x|}{\theta_0})}{\phi(1)}, & \hbox{if $|x|\leq\theta_0$,} \\
\frac{\theta_0}{|x|}\,\frac{\phi(1)}{\phi(\frac{|x|}{\theta_0})}, & \hbox{otherwise.}
\end{array}
\right.
$$
In Figure \ref{scalefig} we give the resulting function for $\theta_0=1$.
\begin{figure}[h]
 \centering
  \includegraphics[height=2in,width=2.5in]{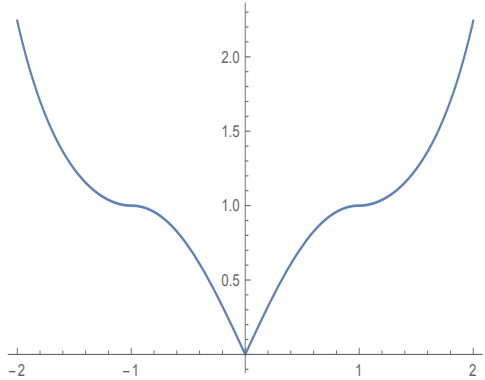}\\
  \caption{The function $V_{npar}$ for the scale family $f(x|\theta)=\phi(x/\theta)/\theta$ and $\theta_0=1$.}\label{scalefig}
\end{figure}
}

\end{exam}

\section{Parametric empirical Bayes}\label{parametric}
We will pursue the parametric empirical Bayes approach for evidence that is normally distributed. The measurement
$X$ has a ${\cal N}(\theta,\sigma^2)$ distribution, thus adhering to the blood alcohol example. So, with $\phi$ denoting the standard normal density we have
\begin{equation}\label{normal}
f(x|\theta)=\frac{1}{\sigma}\phi\Big(\frac{x-\theta}{\sigma}\Big).
\end{equation}
Let us now assume that the prior density is also normal, more specifically ${\cal N}(\mu, \tau^2)$.
This prior density is equal to
\begin{equation}\label{pardensity}
\pi_{\mu,\tau^2}(\theta)=\frac{1}{\tau}\phi\Big(\frac{\theta-\mu}{\tau}\Big).
\end{equation}

In the previous section, in the nonparametric approach, we have put no other restrictions on the prior $\pi$ but the prior probabilities of the judge. These in fact impose a fixed quantile value for $\pi$ at $\theta_0$.

We will consider three situations. First we assume that the prior normal density and its parameters are known.
In fact this means that we assume that the judge chooses his or her personal normal prior and implicitly the prior probabilities (\ref{restrictions}).
This would probably require a  not realistic knowledge of statistics from the judge but the computations are crucial for the next two situations.
Next we return to our original setting where the judge only provides the prior probabilities on $H_p$ and $H_d$, thus setting a restriction on the
parameters $\mu$ and $\tau^2$. Under this restriction we will the estimate the parameters by maximum likelihood and derive the resulting strength of evidence.
This will be done first for the case where the judge is a priori balanced, i.e. the prior odds are equal to one. We will also consider the much more complicated case where
the prior probability of $H_p$ is larger than the prior probability of $H_d$, i.e. prior odds larger than one. The other case can be treated similarly but seems less realistic.

\subsection{Strength of evidence for a known normal prior}

If we assume a completely known normal distribution ${\cal N}(\mu, \tau^2)$ as prior then we can derive the following theorem.

\begin{thm}
Assume that the evidence has a ${\cal N}(0, \sigma^2)$ distribution with $\sigma^2$ known, and that the prior $\pi$ is a known ${\cal N}(\mu, \tau^2)$ distribution.
Then the strength of evidence (\ref{voe}) is  equal to
\begin{equation}\label{voe1}
V_{\mu,\tau^2}(x)={
{\left(1-
\Phi\left(\frac{\theta_0- \frac{\tau^2x+\sigma^2\mu}{\sigma^2+\tau^2} }{\frac{\sigma\tau}{\sqrt{\sigma^2+\tau^2}}}\right)
\right)}\over{\Phi\left(\frac{\theta_0-\frac{\tau^2x+\sigma^2\mu}{\sigma^2+\tau^2} }{\frac{\sigma\tau}{\sqrt{\sigma^2+\tau^2}}}\right)}}
\frac{P(H_d)}{P(H_p)}
=\frac{\Lambda\left(\frac{ \frac{\tau^2x+\sigma^2\mu}{\sigma^2+\tau^2}-\theta_0}{\frac{\sigma\tau}{\sqrt{\sigma^2+\tau^2}}}\right)}{
\Lambda\left( \frac{ \mu -\theta_0 }{\tau}\right)} ,
\end{equation}
if $\tau\not=0$, where $\Phi$ denotes the standard normal distribution function and the function $\Lambda$ is defined by
\begin {equation}\label{Lambda}
\Lambda(y)=\frac{1-\Phi(-y)}{\Phi(-y)}=\frac{\Phi( y)}{\Phi(-y)}
\end{equation}
For $\tau=0$ we get $V_{\mu,\tau^2}(x)=1$, i.e. the limit for $\tau\to 0$ of (\ref{voe1}).
\end{thm}

\begin{figure}[h]
 \centering
  \includegraphics[height=2in,width=2.5in]{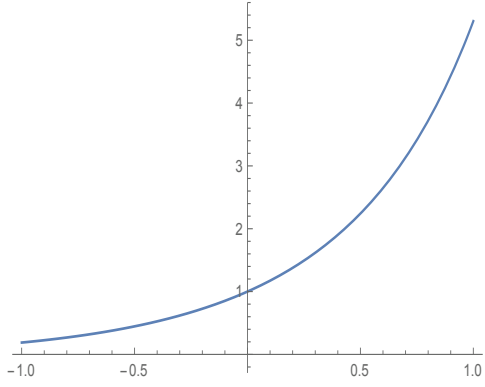}\\
  \caption{The function $\Lambda$.}\label{Lambdafig}
\end{figure}

\noindent{\em Proof}

Similar to (\ref{g}) the density of the evidence is now given by
\begin{eqnarray}
g_{\mu,\tau}(x)&=&\int_{-\infty}^\infty f(x|\theta)\pi_{\mu,\tau^2}(\theta)d\theta\nonumber\\
&=&\int_{-\infty}^\infty
\frac{1}{\sigma}\phi\Big(\frac{x-\theta}{\sigma}\Big)
\frac{1}{\tau}\phi\Big(\frac{\theta-\mu}{\tau}\Big)d\theta\label{gmutau}\\
&=&
\frac{1}{\sqrt{\sigma^2+\tau^2}}\phi\Big(\frac{x-\mu}{\sqrt{\sigma^2+\tau^2}}\Big),\nonumber
\end{eqnarray}
the density of the ${\cal N}(\mu,\sqrt{\sigma^2+\tau^2})$ distribution.

By some  calculations similar to the usual direct proof that the convolution of two normal densities is again normal, we get
$$
\int_{-\infty}^{\theta_0}f(x|\theta)\pi(\theta)d\theta=
\frac{1}{\sqrt{2\pi}\sqrt{\sigma^2+\tau^2}}\,
e^{-\frac{1}{2}\,\frac{(x-\mu)^2}{\sigma^2+\tau^2}}
\Phi\Big(\frac{\theta_0-\frac{\tau^2x+\sigma^2\mu}{\sigma^2+\tau^2}}{\frac{\sigma\tau}{\sqrt{\sigma^2+\tau^2}}}\Big)
$$
and
$$
\int_{\theta_0}^{\infty}f(x|\theta)\pi(\theta)d\theta=
\frac{1}{\sqrt{2\pi}\sqrt{\sigma^2+\tau^2}}\,
e^{-\frac{1}{2}\,\frac{(x-\mu)^2}{\sigma^2+\tau^2}}\Big(1-
\Phi\Big(\frac{\theta_0-\frac{\tau^2x+\sigma^2\mu}{\sigma^2+\tau^2}}{\frac{\sigma\tau}{\sqrt{\sigma^2+\tau^2}}}\Big)
\Big)
$$
Substituting these values in (\ref{voe}), noting $P(H_p)=1-\Phi\Big(\frac{\theta_0-\mu}{\tau}\Big)$ and $P(H_d)=\Phi\Big(\frac{\theta_0-\mu}{\tau}\Big)$ yields (\ref{voe1}).

\hfill$\Box$

Note that the argument of $\Lambda$ in (\ref{voe1}) can be rewritten into
$$
\frac{\tau^2(x-\theta_0) +\sigma^2(\mu-\theta_0)}{\sigma\tau\sqrt{\sigma^2+\tau^2}},
$$
showing that the strength of evidence is monotone in $x$.

\begin{figure}[h]
 \centering
  \includegraphics[height=2in,width=2.5in]{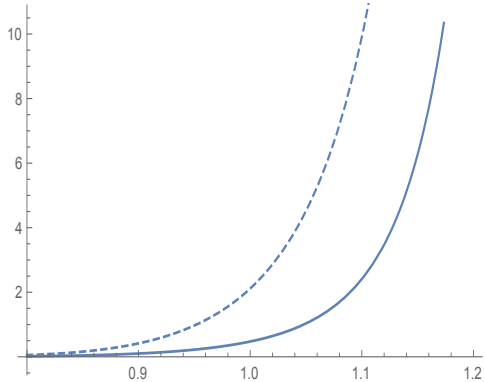}\\
  \caption{Strength of evidence functions (\ref{voe1}) for $\theta_0=1, \sigma=0.1$. Solid: $\mu=1.2, \tau=0.3$ and hence $P(H_p)=0.75$, Dashed: $\mu=0.8, \tau=0.3$ and hence $P(H_p)=0.25$.}\label{mutauknownfig}
\end{figure}

Two examples of the strength of evidence functions are given in Figure \ref{mutauknownfig}.

\subsection{The parametric quantile restriction}

The fixed quantile restriction for a ${\cal N}(\mu, \tau^2)$ prior, in terms of the parameters $\mu$ and $\tau^2$, gives
\begin{equation}\label{parrestriction}
P(H_p)=P(\Theta\geq \theta_0)=1-\Phi\Big(\frac{\theta_0-\mu}{\tau}\Big),
\end{equation}
where $\Theta$ denotes a random variable with density (\ref{pardensity}). We will denote $P(H_p)$ by $\alpha$ and $P(H_d)$ by $1-\alpha$.   Writing $\theta_{u}=\Phi^{-1}(u)$ for the $u$-th quantile we get
\begin{equation}\label{mutau}
0\leq\tau=\frac  {\theta_0-\mu}{\theta_{1-\alpha}}\quad\mbox{and}\quad \mu=\theta_0-\theta_{1-\alpha}\tau.
\end{equation}
We will assume $\alpha\geq 0.5$, so the judge is a priori more convinced of the guilt of the suspect than of his or her innocence.
In (\ref{mutau}) we have to assume that $\alpha>0.5$ and so $\theta_{1-\alpha}\not=0$. If $\alpha=0.5$ then the value of $\tau$ is not determined by the quantile restriction and the restriction only states $\mu=\theta_0$.
Note that $\mu$ is   larger than $\theta_0$ if $\alpha$ is  larger than 0.5.

\subsubsection{The judge is a priori balanced}
Let us first consider the case where the judge is a priori balanced.
We get the following expression for the strength of evidence.

\begin{thm}
Assume that the evidence has a ${\cal N}(\theta, \sigma^2)$ distribution with $\sigma^2$ known, and that the prior $\pi$ is a ${\cal N}(\mu, \tau^2)$ distribution.
Further assume that $P(H_p)=P(H_d)=0.5$. Then the strength of evidence (\ref{voe}), with $\mu=\theta_0$ and $\tau$ estimated by maximum likelihood, is  equal to
\begin{equation}\label{voebalanced}
V(x)=
\Big\{
\begin{array}{ll}
1&, \mbox{if}\quad |x-\theta_0|\leq \sigma,\\
\Lambda(\frac{1}{\sigma}\, \sqrt{(x-\theta_0)^2-\sigma^2}\,\sign(x-\theta_0))&, \mbox{if}\quad |x-\theta_0|\geq \sigma. \\
\end{array}
\end{equation}
\end{thm}

\newpage

\noindent{\em Proof}

The strength of evidence (\ref{voe1}) in this case, with $\alpha=P(H_p)=P(H_d)=0.5$ and hence $\mu=\theta_0$, equals
\begin{equation}\label{voe2}
V(x)=\Lambda\left(\frac{ \frac{\tau^2x+\sigma^2\theta_0}{\sigma^2+\tau^2}-\theta_0}{\frac{\sigma\tau}{\sqrt{\sigma^2+\tau^2}}}\right)
=\Lambda\left((x-\theta_0 )\frac{\tau}{\sigma\sqrt{\sigma^2+\tau^2}}\right)
\end{equation}
where the parameter  $\tau$ still has to be determined.

Now, applying maximum likelihood to estimate $\tau$, we want to maximize $g_{\theta_0,\tau}(x)$, for a fixed observation $x$ and given $\sigma^2$, over the parameter    $\tau$ of the prior density under the restriction (\ref{parrestriction}).

 The prior density (\ref{pardensity}) equals
$$
\pi(\theta)=\frac{1}{\tau \sqrt{2\pi}}\,e^{-\frac{1}{2}\frac{(\theta -\theta_0)^2}{\tau^2}}.
$$
The density $g$ of the evidence is now equal to
$$
\frac{1}{\sqrt{\sigma^2+\tau^2}\, \sqrt{2\pi}}\,e^{-\frac{1}{2}\frac{(x -\theta_0)^2}{\sigma^2+\tau^2}},
$$
the density of the ${\cal N}(\theta_0,\sigma^2+\tau^2)$ distribution.
We will choose $\tau$ such that this expression, for fixed $x$, $\sigma^2$ and $\theta_0$ is maximized. Thus we apply the likelihood principle in choosing $\tau$.
The derivative with respect to $\tau$ of the function
$$
\frac{1}{\sqrt{\sigma^2+\tau^2}}\, \frac{1}{\sqrt{2\pi}}\,e^{-\frac{1}{2}\frac{(x -\theta_0)^2}{\sigma^2+\tau^2}}
=\frac{1}{\sqrt{\sigma^2+\tau^2}}\,\phi\Big(\frac{x-\theta_0}{\sqrt{\sigma^2+\tau^2}}\Big)
$$
is equal to
$$
\tau\,\Big((x-\theta_0)^2-\sigma^2-\tau^2 \Big)
\frac{1}{(\sigma^2+\tau^2)^{5/2}}\,\phi\Big(\frac{x-\theta_0}{\sqrt{\sigma^2+\tau^2}}\Big).
$$
This function is negative for all positive $\tau$ if $|x-\theta_0|\leq \sigma$. Otherwise it equals zero at $\tau=\sqrt{(x-\theta_0)^2-\sigma^2}$. This shows that the maximizing   non negative $\tau(x)$ is given by
\begin{equation}
\tau(x)=\Big\{
\begin{array}{ll}
0&, \mbox{if}\quad |x-\theta_0|\leq \sigma,\\
\sqrt{(x-\theta_0)^2-\sigma^2}&, \mbox{if}\quad |x-\theta_0|\geq \sigma. \\
\end{array}
\end{equation}
The strength of evidence is then equal to
$$
V(x)=\Lambda\left((x-\theta_0 )\frac{\tau(x)}{\sigma\sqrt{\sigma^2+\tau(x)^2}}
\right)\,.
$$
After rewriting this expression we get (\ref{voebalanced}).

\hfill$\Box$

\begin{exam}{Blood alcohol measurements continued.}

{\rm In Figure \ref{Combinedfig} we have plotted the nonparametric and parametric balanced strength of evidence functions for
the blood alcohol example.
\begin{figure}[h]
 \centering
  \includegraphics[height=3in,width=3.5in]{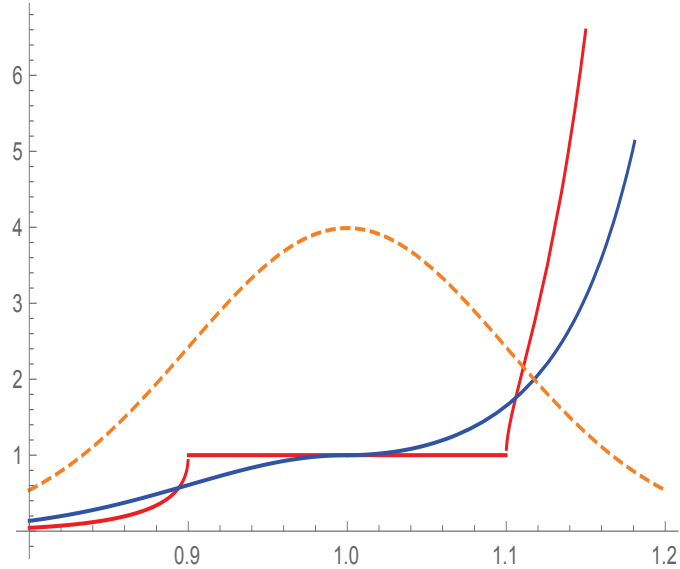}\\
  \caption{Blue: nonparametric, Red: parametric balanced, Orange: density of $X$. Here $\theta_0=1$ and $\sigma=0.1$.}\label{Combinedfig}
\end{figure}

On the flat part the strength of evidence in the parametric setting is equal to one which means that the evidence does not change the prior belief of the judge.
}
\end{exam}
\subsubsection{The unbalanced case: unequal a priori probabilities}

Let us consider the case where the judge is not a priori balanced. This means that $\alpha =P(H_p)\not= 0.5$. We will assume the more realistic case where $\alpha>0.5$. So we assume the judge is
a priori more convinced in $H_p$ than in $H_d$. The next theorem gives some properties of the resulting strength of evidence function. Its proof is given in the appendix.
\begin{thm}\label{unequalthm}
Assume that the evidence has a ${\cal N}(\theta, \sigma^2)$ distribution with $\sigma^2$ known, and that the prior $\pi$ is a ${\cal N}(\mu, \tau^2)$ distribution.
Further assume that $\alpha=P(H_p)>0.5$. Then the strength of evidence (\ref{voe}), with $\mu$ and $\tau$ estimated by maximum likelihood is  equal to one on an interval $[x_0,\theta_0]$ with $x_0<\theta_0$ .
\end{thm}

\begin{exam}{Blood alcohol measurements continued.}

{\rm In Figure \ref{Combinedfig2} we have plotted the nonparametric and parametric  strength of evidence functions with different a priori probabilities for
the blood alcohol example.

\begin{figure}[h]
  \centering
  \includegraphics[height=3in,width=3.5in]{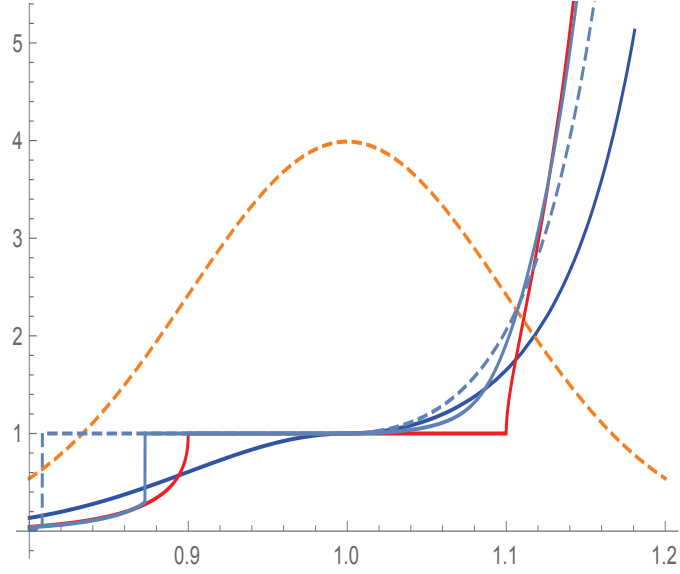}\\
  \caption{Blue: nonparametric, Red: parametric balanced, Grey: $\alpha=0.55$, Grey dashed: $\alpha=0.75$, Orange: density of $X$. Here $\theta_0=1, \sigma=0.1$.}\label{Combinedfig2}
\end{figure}
}
\end{exam}

We observe from the pictures  that the strength of evidence function is strictly increasing before $x_0$ and after $\theta_0$. We also observe a jump at the left end point of the flat part, Further numerical computations have confirmed this.

\section{Conclusions}
In Section \ref{nonparametric} we show that the ratio of two suprema expression for the strength of evidence (\ref{supsratio2}) in the context
of composite hypotheses can be interpreted as an estimate of the strength of evidence if we adopt a nonparametric empirical
Bayes approach with a quantile restriction on the prior. We estimate the nonparametric prior by nonparametric maximum likelihood.
This provides an alternative interpretation.

If we restrict this approach to parametric empirical Bayes with a normal prior we get a flat part in the strength of evidence function. In the case that the judge
is a priori in favour of the prosecutors hypothesis   we also see a jump at the left end point of the flat part. These parametric empirical Bayes results are interesting but only from a mathematical
point of view. In practice they are of limited use.

\section{Appendix}

\subsection{Proof of Theorem \ref{unequalthm}}

From the relations (\ref{mutau}) imposed by the quantile restriction, writing $\mu$ in terms of $\tau$ as in (\ref{mutau}), we see that the density of the evidence (\ref{gmutau}) equals
$$
g_\tau (x)=\frac{1}{\sqrt{\sigma^2+\tau^2}}\,\frac{1}{\sqrt{2\pi}}
e^{-\frac{1}{2}\frac{(x -\theta_0+\theta_{1-\alpha}\tau)^2}{\sigma^2+\tau^2}}=
\frac{1}{\sqrt{\sigma^2+\tau^2}}\,\phi
\Big(\frac{x -\theta_0+\theta_{1-\alpha}\tau)}{\sqrt{\sigma^2+\tau^2}}\Big).
$$
We will determine $\tau$ by  maximum likelihood, i.e. by maximizing this likelihood function over $\tau\geq 0$,  for the evidence $x$ fixed and   $\sigma^2$ known.
To achieve this we analyse its derivative with respect to $\tau$.

The derivative with respect to $\tau$ is equal to
$$
\frac{\partial}{\partial \tau}\, g_\tau (x)=P(\tau)\,\frac{1}{(\sigma^2+\tau^2)^{5/2}}\,\phi
\Big(\frac{x-\theta_0+\theta_{1-\alpha}\tau}{\sqrt{\sigma^2+\tau^2}}\Big) ,
$$
with the third degree polynomial $P$ in $\tau$ defined by
\begin{equation}\label{Ptau}
P(\tau)=-\tau^3+\theta_{1-\alpha}(x-\theta_0)\tau^2+\Big((x-\theta_0)^2-\sigma^2(\theta_{1-\alpha}^2+1)\Big)\tau-\theta_{1-\alpha}(x-\theta_0)\sigma^2.
\end{equation}
Note that the sign of this derivative equals the sign of $P(\tau)$.

The value of $P$ at zero equals $P(0)=-\theta_{1-\alpha}(x-\theta_0)\sigma^2$. Recalling that $\alpha>0.5$, and hence $\theta_{1-\alpha}<0$, this value  is negative for $x<\theta_0$ and positive otherwise.

The derivative of the polynomial $P$ is equal to
\begin{equation}\label{Ptauprime}
P'(\tau)=-3\tau^2+2\theta_{1-\alpha}(x-\theta_0)\tau +\Big((x-\theta_0)^2-\sigma^2(\theta_{1-\alpha}^2+1)\Big) .
\end{equation}
Note that this is a downward opening parabola.
The determinant of this parabola equals $4((\theta_{1-\alpha}^2+3)(x-\theta_0)^2-3(\theta_{1-\alpha}^2+1)\sigma^2)$.
Hence the parabola has no roots if
\begin{equation}\label{twozeros}
(x-\theta_0)^2< 3 \sigma^2\,\frac{ (\theta_{1-\alpha}^2+1)}{\theta_{1-\alpha}^2+3}.
\end{equation}
Hence for these values of $x$, close to $\theta_0$, the parabola  $P'$   is strictly negative.
This implies that $P$ is strictly decreasing for such values.

The second derivative of $P$ is equal to
\begin{equation}\label{Ptaudoubleprime}
P''(\tau)=-6\tau +2\theta_{1-\alpha}(x-\theta_0).
\end{equation}
Its value at $\tau =0$ equals $2\theta_{1-\alpha}(x-\theta_0)$.
This value is positive for $x<\theta_0$ and negative otherwise.

Let us first  consider $x\geq \theta_0$. Then  $P''(0)\leq 0$. Hence $P'$ is decreasing for positive $\tau$.
We have
\begin{equation}
P'(0)=\Big((x-\theta_0)^2-\sigma^2(\theta_{1-\alpha}^2+1)\Big).
\end{equation}
If $(x-\theta_0)^2\leq \sigma^2(\theta_{1-\alpha}^2+1)$ then $P'$ is negative for all positive $\tau$ and $P$ is decreasing for all positive $\tau$.
Otherwise, $P'$ is positive until a certain $\tau$ value. To the right of this value it is negative. Hence $P$ increases at first and then decreases
to minus infinity. Since $P(0)$ is positive if $x\geq \theta_0$, in all considered cases there is a unique strictly positive value of $\tau$ that maximizes
$g_\tau (x)$.

Next consider $x<\theta_0$.
Then  $P''(0)> 0$. Hence $P'$ has its maximum to the right of zero. $P'$ is positive until a certain $\tau$ value. To the right of this value it is negative.
Hence $P$ increases at first and then decreases
to minus infinity.
If $x<\theta_0$ then $P(0)$ is negative and $P$ has a unique maximum at some non negative $\tau$. If this maximum is negative,
for instance if $x$ is close enough to $\theta_0$ to satisfy (\ref{twozeros}), then $P$ is decreasing for
all positive $\tau$ and hence $P$ is negative for all positive $\tau$. In that case the value of $\tau$ that maximizes $g_\tau(x)$ is equal to zero.
If the maximum of $P$ is positive then $g_\tau(x)$ decreases from $g_0(x)$ at zero at first. It then starts increasing to a maximum from which it decreases to zero at infinity.
If this maximum is smaller than $g_0(x)$ then the maximizing value of $\tau$ still equals zero. If the maximum is larger than $g_0(x)$ then the positive
$\tau$ value, at which the maximum is attained, is the maximizing value of $g_\tau(x)$.

\hfill$\Box$

\section*{Acknowledgement} I would like to thank Marjan Sjerps for inspiring this research an her remarks on a previous version of the paper.

\end{document}